\documentclass[sigconf,submission]{acmart}
\usepackage{xcolor, soul}
\sethlcolor{green}
\def\BibTeX{{\rm B\kern-.05em{\sc i\kern-.025em b}\kern-.08emT\kern-.1667em\lower.7ex\hbox{E}\kern-.125emX}}
    
\DeclareMathOperator{\diag}{diag}

\begin{document}

%

\title{Anti-Money Laundering in Bitcoin: Experimenting with Graph Convolutional Networks for Financial Forensics}
\renewcommand{\shorttitle}{Anti-Money Laundering in Bitcoin}



\author{Mark Weber}
\authornote{Both authors contributed equally to this research.}
\affiliation{%
  \institution{MIT-IBM Watson AI Lab}}
\email{mrweber@mit.edu}

\author{Giacomo Domeniconi}
\authornotemark[1]
\affiliation{%
  \institution{IBM Research}}
\email{Giacomo.Domeniconi1@ibm.com}

\author{Jie Chen}
\affiliation{%
  \institution{MIT-IBM Watson AI Lab}}
\email{chenjie@us.ibm.com}

\author{Daniel Karl I. Weidele}
\affiliation{%
  \institution{IBM Research AI}}
\email{daniel.karl@ibm.com}

\author{Claudio Bellei}
\affiliation{%
  \institution{Elliptic}}
\email{claudio@elliptic.co}

\author{Tom Robinson}
\affiliation{%
  \institution{Elliptic}}
\email{tom@elliptic.co}

\author{Charles E. Leiserson}
\affiliation{%
  \institution{MIT CSAIL}}
\email{cel@mit.edu}

\renewcommand{\shortauthors}{Weber and Domeniconi, et al.}

\begin{abstract}
Anti-money laundering (AML) regulations play a critical role in safeguarding financial systems, but bear high costs for institutions and drive financial exclusion for those on the socioeconomic and international margins. The advent of cryptocurrency has introduced an intriguing paradox: pseudonymity allows criminals to hide in plain sight, but open data gives more power to investigators and enables the crowdsourcing of forensic analysis. Meanwhile advances in learning algorithms show great promise for the AML toolkit. In this workshop tutorial, we motivate the opportunity to reconcile the cause of safety with that of financial inclusion. We contribute the Elliptic Data Set, a time series graph of over 200K Bitcoin transactions (nodes), 234K directed payment flows (edges), and 166 node features, including ones based on non-public data; to our knowledge, this is the largest labelled transaction data set publicly available in any cryptocurrency. We share results from a binary classification task predicting illicit transactions using variations of Logistic Regression (LR), Random Forest (RF), Multilayer Perceptrons (MLP), and Graph Convolutional Networks (GCN), with GCN being of special interest as an emergent new method for capturing relational information. The results show the superiority of Random Forest (RF), but also invite algorithmic work to combine the respective powers of RF and graph methods. Lastly, we consider visualization for analysis and explainability, which is difficult given the size and dynamism of real-world transaction graphs, and we offer a simple prototype capable of navigating the graph and observing model performance on illicit activity over time. With this tutorial and data set, we hope to a) invite feedback in support of our ongoing inquiry, and b) inspire others to work on this societally important challenge.

\end{abstract}

%


%

\maketitle

\section{Toward Financial Inclusion}
``It's expensive to be poor.'' This is a common credo among advocates for financial inclusion. It speaks to the fact that those on the margins of society suffer from restricted access to the financial system and higher relative costs of participation. 

The problem of restricted access (e.g. the ability to sign up for a bank account) is, in part, an unintended consequence of increasingly stringent anti-money laundering (AML) regulations, which, while essential for safeguarding the financial system, have a disproportionately negative effect on low-income people, immigrants, and refugees \citep{WorldBank2012}. Approximately 1.7 billion adults are unbanked \citep{WorldBank2017}. The problem of higher relative costs is also, in part, a function of AML policy, which enforces high fixed costs of compliance on money service businesses (MSBs) along with the fear of criminal and monetary penalties for noncompliance -- ``low value" customers just aren't worth the risk. Consider global remittances to low-and-middle-income countries, which reached a record high \$529 billion in 2018, far outpacing the global aid contribution of \$153 billion. The current average cost of sending \$200 is an expensive 7 percent, with some countries suffering rates of over 10 percent. The United Nations Sustainable Development Goal number 10.7 targets a reduction to 3 percent by 2030.\citep{WorldBank2019}

And yet AML regulations cannot be summarily dismissed as over burdensome. Multi-billion dollar illicit industries like drug cartels, human trafficking, and terrorist organizations cause intense human suffering around the world. The recent 1Malaysia Development Berhad (1MDB) money laundering scandal robbed the Malaysian people of over \$11 billion in taxpayer funds earmarked for the nation's development \citep{1MDB}, with mega-fines and criminal indictments for Goldman Sachs among others implicated in the wrongdoing. The even more recent Danske Bank money laundering scandal in Estonia, which served as a hub for an estimated \$200 billion in illicit money flows from Russia and Azerbaijan, similarly extracted an incalculable toll on innocent citizens of these countries and served implicated institutions like Danske Bank and Deutsche Bank with billions of dollars in losses \citep{Danske}.

Money laundering is not a victimless crime, and current methods for the traditional financial system are doing a poor job of stopping it. Without reducing this complex challenge to data analysis alone, we pose the question: \textit{with the right tools and open data, can we help reconcile the need for safety with the cause of financial inclusion?}

\subsection{AML in a cryptocurrency world}

The advent of cryptocurrency introduced by Bitcoin \cite{nakamoto} ignited an explosion of technological and entrepreneurial interest in payment processing. Around the world, money transfer startups spun up to compete with legacy banks and MSBs like Western Union. They focused on enabling low-cost, peer-to-peer transfers of cash within and across borders using Bitcoin and other cryptocurrencies as the ``rails'' (a commonly used term in this space). Many explicitly targeted remittances and championed the cause of financial inclusion. Alongside these entrepreneurs grew a community of academics and policy advocates supporting updated regulatory considerations for cryptocurrency.

Dampening this excitement was Bitcoin's bad reputation. Many criminals used Bitcoin's pseudonymity to hide in plain sight, conducting ransomware attacks and operating dark marketplaces for the exchange of illegal goods and services.

In May 2019, the Financial Crimes Enforcement Network (FinCEN) of the United States issued new guidance on how the Bank Secrecy Act (BSA) of 1970 applies to cryptocurrency, or what FinCEN calls convertible virtual currencies (CVC) \cite{FinCEN}. Consistent with the BSA, the guidance calls for MSBs to generate individualized risk assessments measuring exposure to money laundering, terrorism finance, and other financial crime. These assessments are based on customer composition, geographies served, and financial products or services offered. The assessments must inform the management of customer relationships, including the implementation of controls commensurate with risk; in other words, MSBs must not only report suspicious accounts, but must also take action against them (e.g. freeze them or shut them down). The guidance defines a ``well-developed risk assessment'' as ``assisting MSBs in identifying and providing a comprehensive analysis of their individual risk profile.'' Reinforcing the Know Your Customer (KYC) requirements of the BSA, the guidance requires MSBs to ``know enough about their customers to be able to determine the risk level they represent to the institution.''

What it means to ``know enough'' about one's customer is the subject of much debate in compliance and policy circles. In practice, one of the most challenging aspects of this is an implicit but effectively enforced requirement to not only know your customer, but to know your \textit{customer's} customer. In the fragmented data ecosystem of traditional finance, this aspect of compliance is often executed by phone calls between MSBs. But in the open system of Bitcoin, the full graph transaction network data is publicly available, albeit in pseudonymous and unlabelled form.

To meet the opportunity this public data presents, cryptocurrency intelligence companies have emerged to provide AML solutions tailored to the cryptocurrency domain. Whereas the pseudonymity of Bitcoin is an advantage for criminals, the public availability of data is a key advantage for investigators.

\section{The Elliptic Data Set}
\label{dataset}

Elliptic is a cryptocurrency intelligence company focused on safeguarding cryptocurrency ecosystems from criminal activity. For this tutorial and as a contribution to the research community, we present \textit{The Elliptic Data Set}, a graph network of Bitcoin transactions with handcrafted features. As a contribution to the research and AML communities, Elliptic has agreed to share this data set publicly. To our knowledge, it constitutes the world's largest labelled transaction data set publicly available in any cryptocurrency.

\begin{figure*}[h]
  \centering
  \includegraphics[width=.8\linewidth]{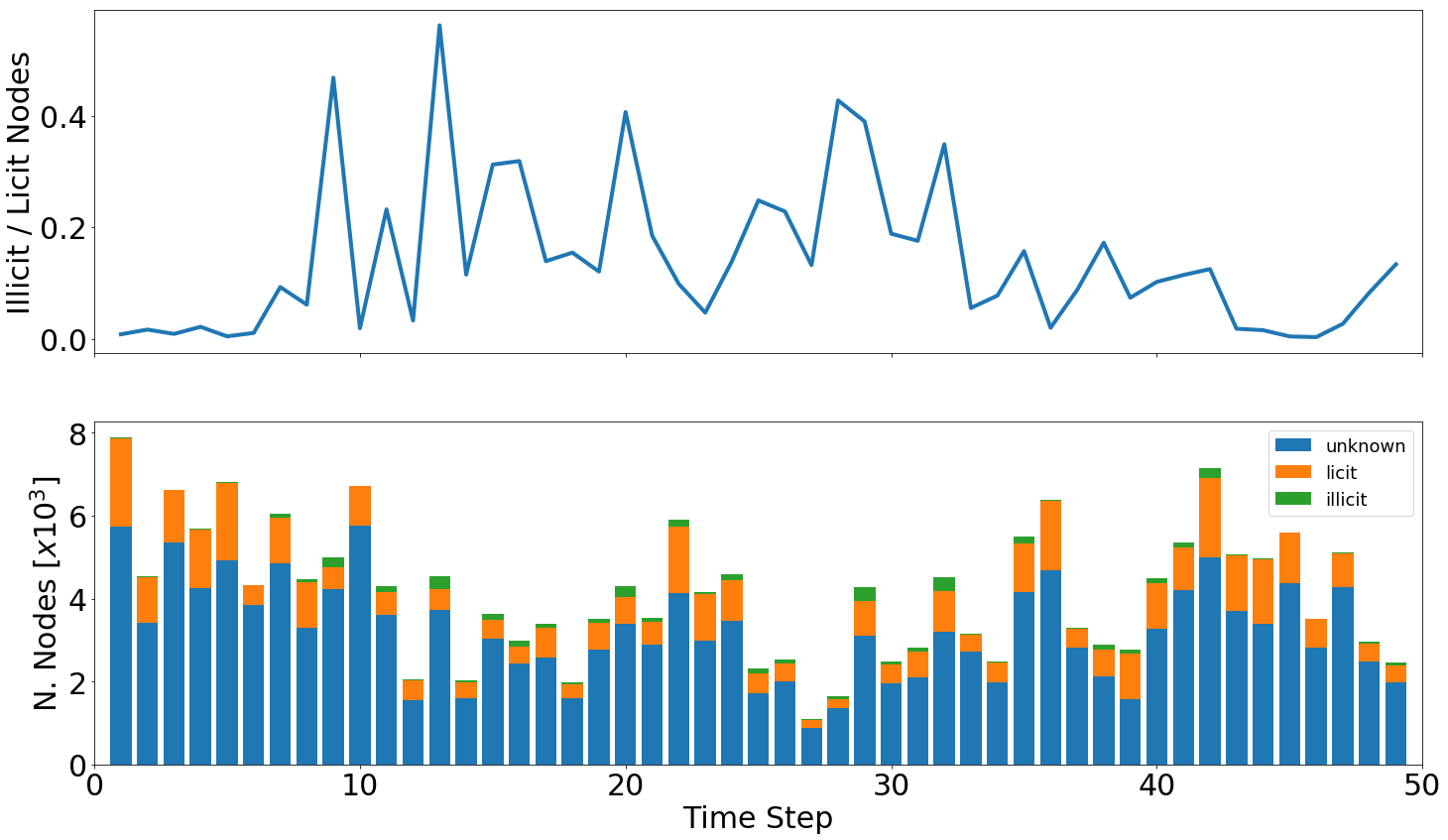}
  \caption{(Top) Fraction of \textit{illicit} vs. \textit{licit} nodes at different time steps in the data set. (Bottom) Number of nodes vs. time step. }
  \label{fig:dataset}
\end{figure*}

\subsection{Graph Construction}

The Elliptic Data Set maps Bitcoin transactions to real entities belonging to \textit{licit} categories (exchanges, wallet providers, miners, licit services, etc.) versus \textit{illicit} ones (scams, malware, terrorist organizations, ransomware, Ponzi schemes, etc.). From the raw Bitcoin data, a graph is constructed and labelled such that the nodes represent transactions and the edges represent the flow of Bitcoin currency (BTC) going from one transaction to the next one. A given transaction is deemed \textit{licit} (versus \textit{illicit}) if the entity initiating the transaction (i.e., the entity controlling the private keys associated with the input addresses of a specific transaction) belongs to a licit (illicit) category\footnote{Note that for simplicity, this argument ignores mixer transactions where the inputs are controlled by multiple entities.}. Importantly, all features are constructed using only publicly available information.

\subsubsection{Nodes and Edges.} There are 203,769 node transactions and 234,355 directed edge payments flows. For perspective, using the same graph representation the full Bitcoin network has approximately 438M nodes and 1.1B edges as of this writing. In the Elliptic Data Set, two percent (4,545) are labelled \textit{class1 (illicit)}. Twenty-one percent (42,019) are labelled \textit{class2 (licit)}. The remaining transactions are not labelled with regard to licit versus illicit, but have other features.

\subsubsection{Features.} Each node has associated 166 features. The first 94 features represent \textit{local information} about the transaction -- including the time step, number of inputs/outputs, transaction fee, output volume and aggregated figures such as average BTC received (spent) by the inputs/outputs and average number of incoming (outgoing) transactions associated with the inputs/outputs. The remaining 72 features, called \textit{aggregated features}, are obtained by aggregating transaction information one-hop backward/forward from the center node - giving the maximum, minimum, standard deviation and correlation coefficients of the neighbour transactions for the same information data (number of inputs/outputs, transaction fee, etc.).

\subsubsection{Temporal Information.} A time stamp is associated with each node, representing an estimate of the time when the transaction is confirmed by the Bitcoin network. There are 49 distinct time steps, evenly spaced with an interval of about two weeks. Each time step contains a single connected component of transactions that appeared on the blockchain within less than three hours between each other; there are no edges connecting different time steps. Clearly the nodes in a specific time step have associated time stamps very close to each other, so effectively each one of them can be thought of as an instantaneous ``snapshot'' in time. The number of nodes for each time step is reasonably uniform over time (ranging from 1,000 to 8,000 nodes). See Figure~\ref{fig:dataset}.


\subsection{Notes on Feature Construction}

The licit versus illicit labelling process is informed by a heuristics-based reasoning process. For example, a higher number of inputs and the reuse of the same address is commonly associated with higher address-clustering \cite{harrigan2016unreasonable}, which results in a degrade of anonymity for the entity signing the transaction. On the other hand, consolidating funds controlled by multiple addresses in one single transaction provides benefits in terms of transaction costs (fee). It follows that entities eschewing anonymity-preserving measures for large volumes of user requests are likely to be licit (e.g. exchanges). In contrast, illicit activity may tend to favor transactions with a lower number of inputs to reduce the impact of de-anonymizing address-clustering techniques.


Additionally, there are two major challenges in building features for Bitcoin transactions. The first is rooted in the size of the Bitcoin blockchain amounting to 200GB of compressed data and about 400 million addressed transactions. Though not all transactions are included in the subset used in this study, it is still necessary to access the complete blockchain in order to observe the full history of wallets participating in the selected transactions. To overcome this, Elliptic uses a high-performance all-in-memory graph engine for the computation of features.

The second challenge arises from the underlying graph structure of the data and the heterogeneity in the number of neighbors a transaction can have. In building the 72 aggregated features, the problem of heterogeneous neighborhoods is addressed by naively constructing statistical aggregates (minimum, maximum, etc.) of the local features of a neighbor transaction. In general, this solution is sub-optimal because it carries a significant loss of information. We address this in our forthcoming discussion of graph deep learning methods, which may better account for the local graph topology.

\section{Task and Methods}
At a high level, AML analytics is an anomaly detection challenge of accurately classifying a small number of illicit transactions in massive, ever-growing data sets. Industry standard high false positive rates of upwards of 90\% inhibit this effort. We want to reduce false positive rates without increasing false negative rates, i.e. include more innocent people without allowing more criminals. Logistic Regression and Random Forest are among the benchmark methods for this task. Graph deep learning has also emerged as potential tool for AML \cite{Weber}.

In the case of the Elliptic Data Set, the task to be performed on this data is transaction screening for assessing the risk associated with a given transaction to-and-from cryptocurrency wallets. Specifically, each unlabelled Bitcoin transaction is to be classified illicit or licit.

\subsection{Benchmark Methods}
Given the features previously described, benchmark machine learning methods use the first 94 features in supervised learning for binary classification. Such techniques include Logistic Regression \citep{Bishop}, Multilayer Perceptron (MLP) (ibid), and Random Forest \cite{breiman2001random}. In MLP, each input neuron takes in a data feature and the output is a softmax with a probability vector for each class. Logistic Regression and Random Forest are popular for AML, especially when used in concert with one another for their respective advantages---Random Forest for accuracy and Logistic Regression for explainability. These methods, however, do not leverage any graph information.

In the Elliptic Data Set, the local features are enhanced with a set of 72 features that contain information about the immediate neighbourhood. We will see the utilization of these features improves performance. While this approach shows the graph structure carries in the binary classification problem, and that this can be used with standard machine learning techniques, it is challenging to extend the purely feature-based method beyond the immediate neighbourhood. This drawback motivates the use of Graph Convolutional Networks.
    
\subsection{Graph Convolutional Networks (GCN)}\label{sec:gconv}
Deep learning on graph structured data is a subject of rapidly increasing interest ~\citep{Bruna2014,Defferrard2016,Li2016,Hamilton2017,Gilmer2017}. Dealing with combinatorial complexity inherent to graph structures poses scalability challenges for practical applications, and significant strides have been made in addressing these challenges \citep{Kipf2017,Chen2018,Ying2018}. Specifically, we consider Graph Convolutional Networks (GCNs). A GCN consists of multiple layers of graph convolution, which is similar to a perceptron but additionally uses a neighborhood aggregation step motivated by spectral convolution.

Consider the Bitcoin transaction graph from the Elliptic Data Set as $G=(N,E)$, where $N$ is the set of node transactions and $E$ is the set of edges representing the flow of BTC. The $l$-th layer of the GCN takes the adjacency matrix $A$ and the node embedding matrix $H^{(l)}$ as input, and uses a weight matrix $W^{(l)}$ to update the node embedding matrix to $H^{(l+1)}$ as output. Mathematically, we write
\begin{equation}
  H^{(l+1)} =  \sigma(\widehat{A} H^{(l)} W^{(l)}), \label{eqn:gcn}
\end{equation}
where $\widehat{A}$ is a normalization of $A$ defined as:
\[
\widehat{A}=\widetilde{D}^{-\frac{1}{2}}\widetilde{A}\widetilde{D}^{-\frac{1}{2}},
\quad \widetilde{A}=A+I,
\quad \widetilde{D}=\diag\Bigg(\sum_j\widetilde{A}_{ij}\Bigg),
\]
and $\sigma$ is the activation function (typically ReLU) for all but the output layer. The initial embedding matrix comes from the node features; i.e., $H^{(0)}=X$. Let there be $L$ layers of graph convolutions. In the case of node classification, the output layer is the softmax, where $H^{(L)}$ consists of prediction probabilities.

One sees a graph convolution layer is similar to a feed forward layer, except for the multiplication with $\widehat{A}$ in the front. This matrix is motivated by spectral graph filtering on the graph Laplacian matrix and it results from a linear functional of the Laplacian. On the other hand, one may also interpret the multiplication with $\widehat{A}$ as an aggregation of the transformed embeddings of the neighboring nodes.
The parameters of the GCN are the weight matrices $W^{(l)}$, for different layers $l$. 

A 2-layer GCN, as often used, can be neatly written as
\[
H^{(2)}=\text{softmax}(\widehat{A}\cdot\text{ReLU}(\widehat{A}XW^{(0)})\cdot W^{(1)}).
\]
A ``skip'' variant, which we find practically useful, inserts a skip connection between the intermediate embedding $H^{(1)}=\text{ReLU}(\widehat{A}XW^{(0)})$ and the input node features $X$, resulting in the architecture
\[
\widetilde{H}^{(2)}=\text{softmax}(\widehat{A}\cdot\text{ReLU}(\widehat{A}XW^{(0)})\cdot W^{(1)}+X\widetilde{W}^{(1)}),
\]
where $\widetilde{W}^{(1)}$ is a weight matrix for the skip connection.
We call this architecture Skip-GCN. When $W^{(0)}$ and $W^{(1)}$ are zero, Skip-GCN is equivalent to Logistic Regression. Hence, Skip-GCN should be at least as powerful as Logistic Regression.

\subsection{Temporal Modeling}
Financial data are inherently temporal as transactions are time stamped. It is reasonable to assume there exists certain dynamics, albeit hidden, that drive the evolution of the system. A prediction model will be more useful if it is designed in a manner to capture the dynamism. This way, a model trained on a given time period may better generalize to subsequent time steps. The better the model captures system dynamics, which are also evolving, the longer horizon it can forest into.

A temporal model that extends GCN is EvolveGCN~\cite{Pareja2019}, which computes a separate GCN model for each time step. These GCNs are then connected through a recurrent neural network (RNN) to capture the system dynamics. Hence, the GCN model for a future time step is evolved from those in the past, where the evolution captures the dynamism.

In EvolveGCN, the GCN weights are collectively treated as the system state. The model is updated upon an input to the system every time, by using an RNN (e.g., GRU). The input is the graph information at the current time step. The graph information may be instantiated in many ways; in EvolveGCN, it is represented by the embeddings of the top-$k$ influential nodes in the graph.

\section{Experiments}

\begin{table}
  \caption{Illicit classification results. Top part of the table shows results without the leverage of the graph information, for each model are shown results with different input: $AF$ refers to all features, $LF$ refers to the local features, i.e. the first 94, and $NE$ refers to the node embeddings computed by GCN. Bottom part of the table shows results with GCN.}
  \label{tab:results}
  \begin{tabular}{c|ccc|c}
    \toprule
     & \multicolumn{3}{|c|} {Illicit} & MicroAVG \\
    Method &  Precision &  Recall &  $F_1$ &  $F_1$\\
    \midrule
    Logistic Regr$^{AF}$ & 0.404 & 0.593 & 0.481  & 0.931 \\
    Logistic Regr$^{AF+NE}$ & 0.537 & 0.528 & 0.533 & 0.945 \\
    Logistic Regr$^{LF}$ & 0.348 & 0.668 & 0.457 & 0.920 \\
    Logistic Regr$^{LF+NE}$ & 0.518 & 0.571 & 0.543 & 0.945\\
    RandomForest$^{AF}$ & 0.956 & 0.670 & 0.788 & 0.977\\
    RandomForest$^{AF+NE}$ & 0.971 & 0.675 & 0.796 & 0.978 \\
    RandomForest$^{LF}$ & 0.803 & 0.611 & 0.694 & 0.966\\
    RandomForest$^{LF+NE}$ & 0.878 & 0.668 & 0.759 & 0.973 \\
    MLP$^{AF}$ & 0.694 & 0.617 & 0.653 & 0.962 \\
    MLP$^{AF+NE}$ & 0.780 & 0.617 & 0.689 & 0.967 \\
    MLP$^{LF}$ &  0.637 & 0.662 & 0.649 & 0.958 \\
    MLP$^{LF+NE}$ & 0.6819 & 0.5782 & 0.6258  & 0.986 \\
    \midrule
    GCN  & 0.812 & 0.512 & 0.628 & 0.961 \\
    Skip-GCN  & 0.812 & 0.623 & 0.705 & 0.966\\
  \bottomrule
\end{tabular}
\end{table}

Here we show experimental results obtained on the Elliptic Data Set. We performed a 70:30 temporal split of training and test data, respectively. That is, the first 34 time steps are used for training the model and the last 15 for test. We use a temporal split because it reflects the nature of the task. As such, GCN is trained in an inductive setting.

We first tested standard classification models for the licit/illicit prediction using three standard approaches: \textit{Logistic Regression} (with default parameters from the scikit-learn Python package \cite{sklearn_api}), \textit{Random Forest} (also from scikit-learn, with 50 estimators and 50 max features), and \textit{Multilayer Perceptron} (implemented in PyTorch). Our MLP had one hidden layer of 50 neurons and was trained for 200 epochs by using the \textit{Adam} optimizer and a learning rate of 0.001.

\begin{figure*}[t]
  \centering
  \includegraphics[width=\textwidth]{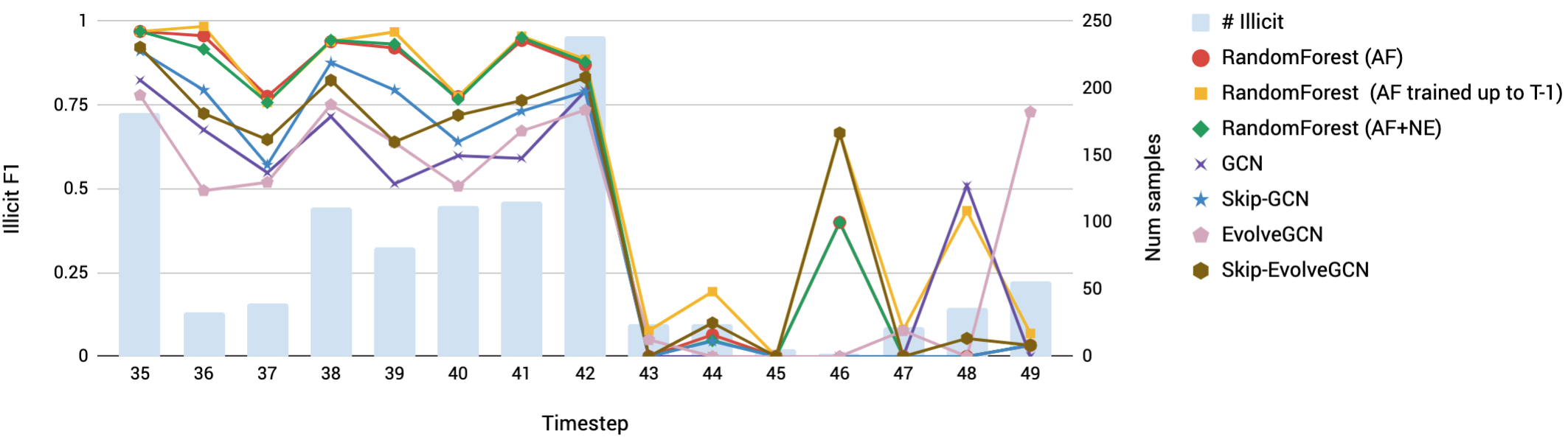} 
  \caption{Illicit $F_1$ results over test time span.}
  \Description{}
  \label{fig:resultovertime}
\end{figure*}

We evaluated these models by using all the 166 features  (referred to as $AF$), as well as only the local ones, i.e., the first 94 (referred to as $LF$). The results are summarized in the top part of Table \ref{tab:results}. 

The bottom part of Table \ref{tab:results} reports the results achieved when we leveraged the graph structure of the data.
We trained the GCN model for 1000 epochs using the \textit{Adam} optimizer with a learning rate of 0.001. In our experiment we used a 2-layer GCN and, after hyper-parameter tuning, we set the size of the node embeddings to be 100.

The task is a binary classification and the two classes are imbalanced (see Figure \ref{fig:dataset}). For AML, more important is the minority class (i.e., the illicit class). 
Hence, we trained the GCN model using a weighted cross entropy loss to provide higher importance to the illicit samples. After hyperparameter tuning, we opted for a $0.3/0.7$ ratio for the licit and illicit classes. 
Table \ref{tab:results} shows the testing results in term of precision, recall, and $F_1$ score for the illicit class. For the sake of completeness, we also show the micro-averaged $F_1$ score. 

Note that GCN and the variant Skip-GCN outperform Logistic Regression, indicating the usefulness of the graph-based method compared to one agnostic to graph information. On the other hand, in this case, the input features are quite informative already.
Using these features alone, Random Forest achieves the best $F_1$ score. The representation power of the input features is also reflected by the gain of Skip-GCN over GCN.

Another insight from Table \ref{tab:results} is obtained from the comparison between methods trained on all the features ($AF$) and those on only the 94 local features ($LF$). For all the three evaluated models, the aggregated information led to higher accuracy, indicating the importance of the graph structure in this context. With this observation, we further evaluated the methods with an \textit{enhanced} input feature set. The goal of this experiment was to show that graph information was useful to enhance the representation of a transaction. In this setting, we concatenated the node embeddings obtained from GCN with the original features $X$. Results show that with the enhanced feature set the accuracy of the model improves, for both full features ($AF+NE$) and local features ($LF+NE$).

Table~\ref{tab:evolvegcn} compares the prediction performance between the non-temporal GCN and the temporal EvolveGCN. EvolveGCN consistently outperforms GCN, although the improvement is not substantial for this data set. One avenue of further investigation is the use of alternative forms of system input to drive the recurrent update inside GRU.

\begin{table}[ht]
\centering
\caption{GCN v.s. EvolveGCN}
\label{tab:evolvegcn}
\begin{tabular}{|c|ccc|ccc|}
\hline
& \multicolumn{3}{c|}{GCN} & \multicolumn{3}{c|}{EvolveGCN}\\
& Precis. & Recall & $F_1$ & Precis. & Recall & $F_1$\\
\hline
Illicit  & 0.812 & 0.623 & 0.705 & 0.850 & 0.624 & 0.720\\
MicroAVG & 0.966 & 0.966 & 0.966 & 0.968 & 0.968 & 0.968\\
\hline
\end{tabular}
\end{table}

\textbf{The Dark Market Shutdown.} 
An important consideration for AML is the robustness of a prediction model with respect to emerging events. One interesting aspect of this data set is the sudden closure of a dark market occurring during the time span of the data (at time step 43). As seen in Figure \ref{fig:resultovertime}, this event causes all methods to perform poorly after the shutdown. Even a Random Forest model re-trained after every test time step, assuming the availability of ground truth after each time, is not able to reliably capture new illicit transactions after the dark market shutdown. The robustness of methods to such events emerges as a major challenge to address.


\section{Discussion}
We have seen Random Forest significantly outperforms Logistic Regression; in fact, it also outperforms GCN even though the latter is empowered by the graph structure information. Random Forest uses a voting mechanism to ensemble the prediction results from a number of decision trees, each trained by using a subsample of the data set. GCN, in contrast, like most deep learning models, uses Logistic Regression as the final output layer; hence, it can be considered a nontrivial generalization of Logistic Regression.

The question arises: \textit{Is it possible to combine a Random Forest with a graph neural network?} One simple idea is to augment the node features with the embeddings computed from GCN before running Random Forest. This idea helps only marginally according to prior experimentation. Another idea, as proposed by~\cite{Kontschieder2015}, is to parameterize every node in the decision tree(s) by using a feed-forward neural network. This idea organically combines Random Forest with neural networks, but it does not suggest how graph information can be incorporated. One possible approach is to replace the Logistic Regression output layer in GCN by this differentiable version of the decision tree, so that end-to-end training is enabled. We leave the execution of this idea as future investigation.



\section{Graph Visualization}
Lastly, in support of analysis and explainability, which are important for AML compliance, we have created a visualization prototype called \emph{Chronograph}. Visualizing a high-dimensional graph imposes a layer of complexity on top of plain feature vectors with respect to explaining model performance. Chronograph aims to address this by supporting the human analyst with an integrated representation of the model.

\subsection{Visual Investigation of the Elliptic Data Set}

In Chronograph, transactions are visualized as nodes of a graph with edges representing the flow of BTC from one transaction to the next. Node coordinates are computed simultaneously across all time steps using the projection technique UMAP \cite{mcinnes2018umap}. This global computation makes layouts comparable across time. The time step slider control at the top of the interface allows the user to navigate through time by rendering only nodes in the selected time step. Illicit transactions are dyed red; licit ones are blue. Unclassified transactions are not colored.

When clicking a transaction node, or entering a transaction ID in the control on the left (filters as substring) the visualization highlights the selected transaction(s) in orange, and all neighboring transactions (in-or-out-flowing) in green. On the left of the interface, the user can see general statistics on the graph and a table about transfer numbers between different transaction classes.

In this simple prototype, Chronograph enables simple exploration scenarios to visually inspect clusters and their existence over time, observe conspicuous transfer patterns, or detect other deviations like single outliers. As a more involved use case we additionally facilitate the degree of freedom of the input to the UMAP computation: raw transaction feature data (Figure 4a), as well as neuron activations of the last layer of the network (Figure 4b) seem to be two interesting alternatives; similar approaches have been proposed for general neural networks \citet{rauber2016visualizing}. Differences in the resulting visualizations would then hint towards peculiarities of the model, i.e. we postulate shifts in similarities among the data can be indicative to explain which underlying features matter to the model.

Figure 4 shows the results from the two alternative inputs for a single time step, with raw feature data in the top and model activations in the bottom row. We further dye the nodes using actual labels in the left column, and GCN-predicted labels in the right column, and obtain a total of 4 network visualizations.

In the model-based layout illicit nodes are less scattered but more concentrated, which seems to be a desirable property: illicit nodes should share some important characteristics, and similarity of nodes yields closer proximity in the layout. However, since they are not perfectly collapsing in one location it is quite plausible there are qualitative differences within the set of illicit nodes. The visualizations further reveals where exactly the model is unable to detect illicit nodes. In case of multiple erroneous predictions in a nearby area this could further hint to a systemic underperformance of the the model. Studying the characteristics of such transactions in detail could inspire the discussion from new angles and lead to further model improvements.

\begin{figure}[h]
  \label{fig:projections}
  \centering
  \includegraphics[width=0.49\linewidth]{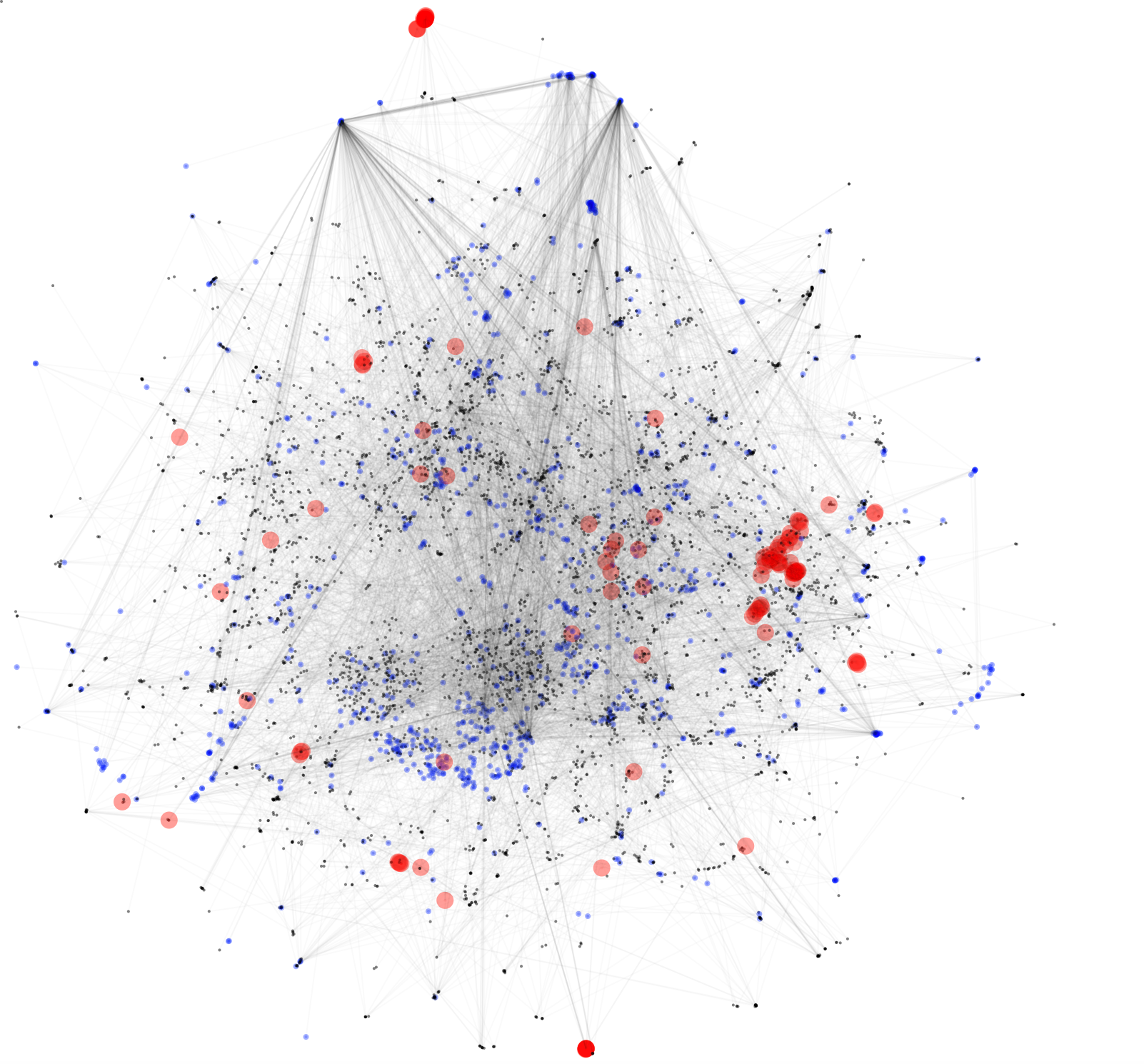}
  \includegraphics[width=0.49\linewidth]{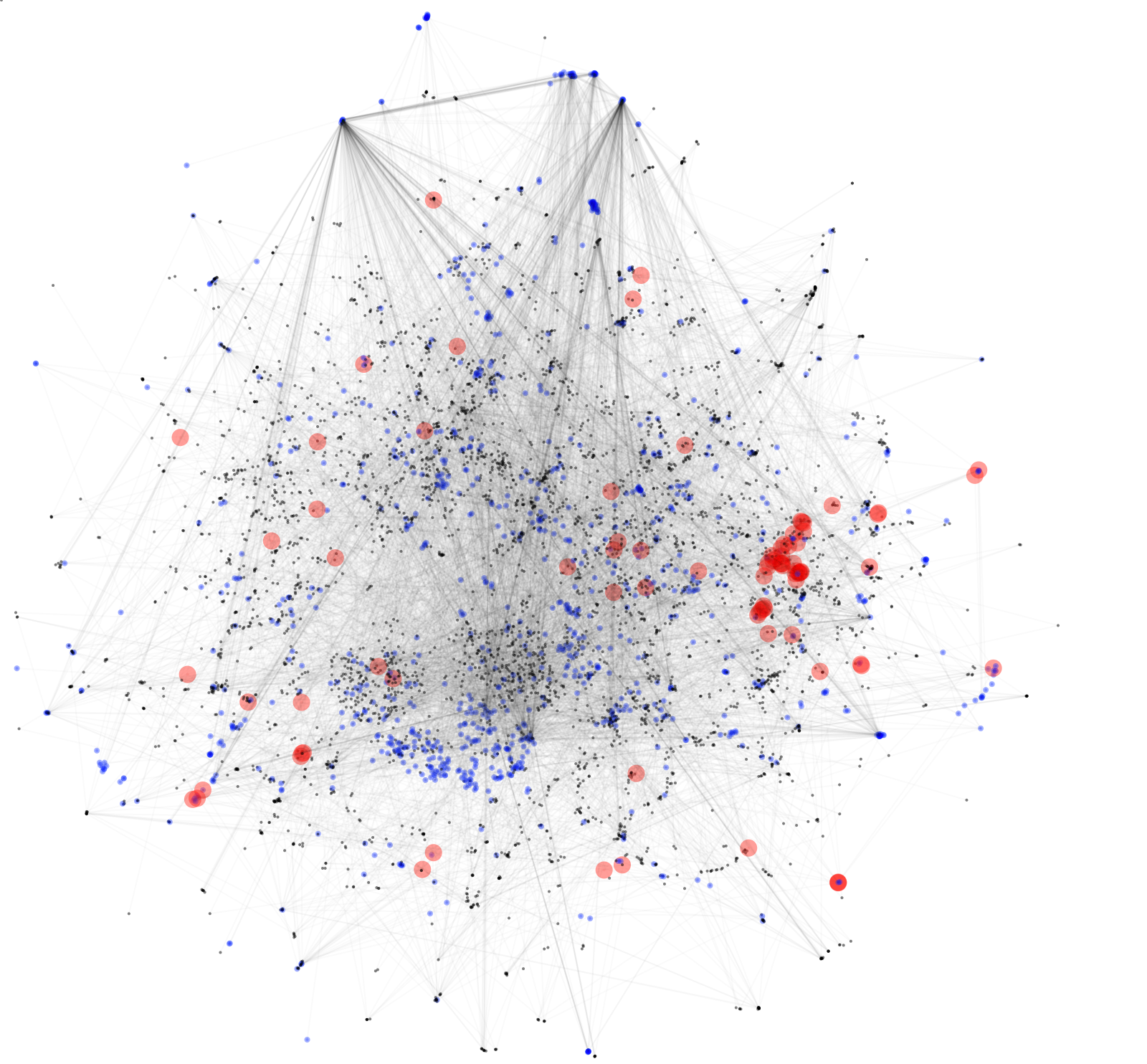}
  \begin{center}
      a) Projection of raw transaction feature vectors
  \end{center}
  \includegraphics[width=0.49\linewidth]{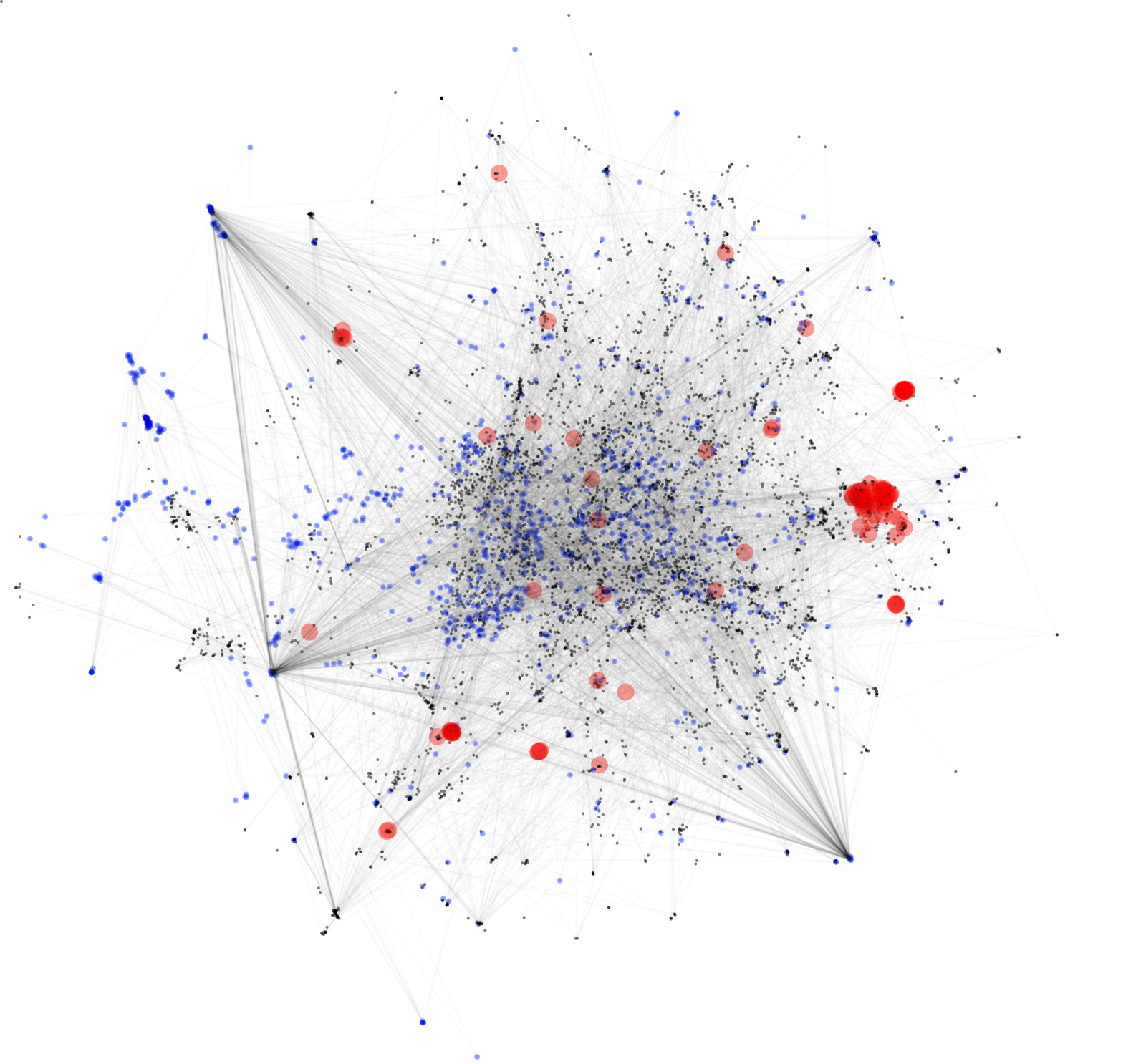}
  \includegraphics[width=0.49\linewidth]{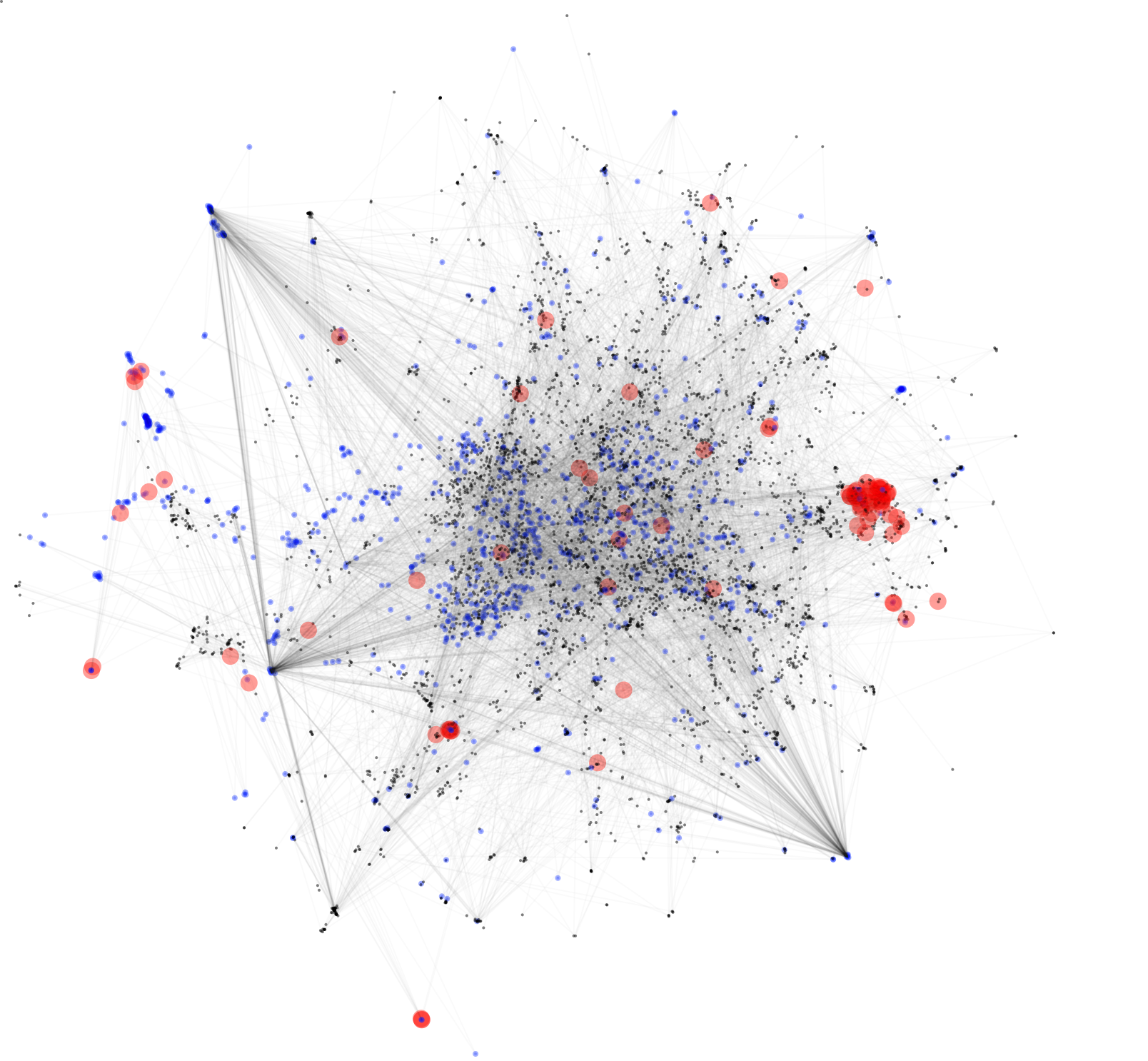}
  \begin{center}
      b) Projection of last GCN layer activations
  \end{center}
  \caption{Two alternative inputs to UMAP projection. Left: colored by input labels; right: colored by GCN prediction.}
  \Description{}
\end{figure}

\begin{figure}[h]
  \label{fig:chronograph}
  \centering
  \includegraphics[width=\linewidth]{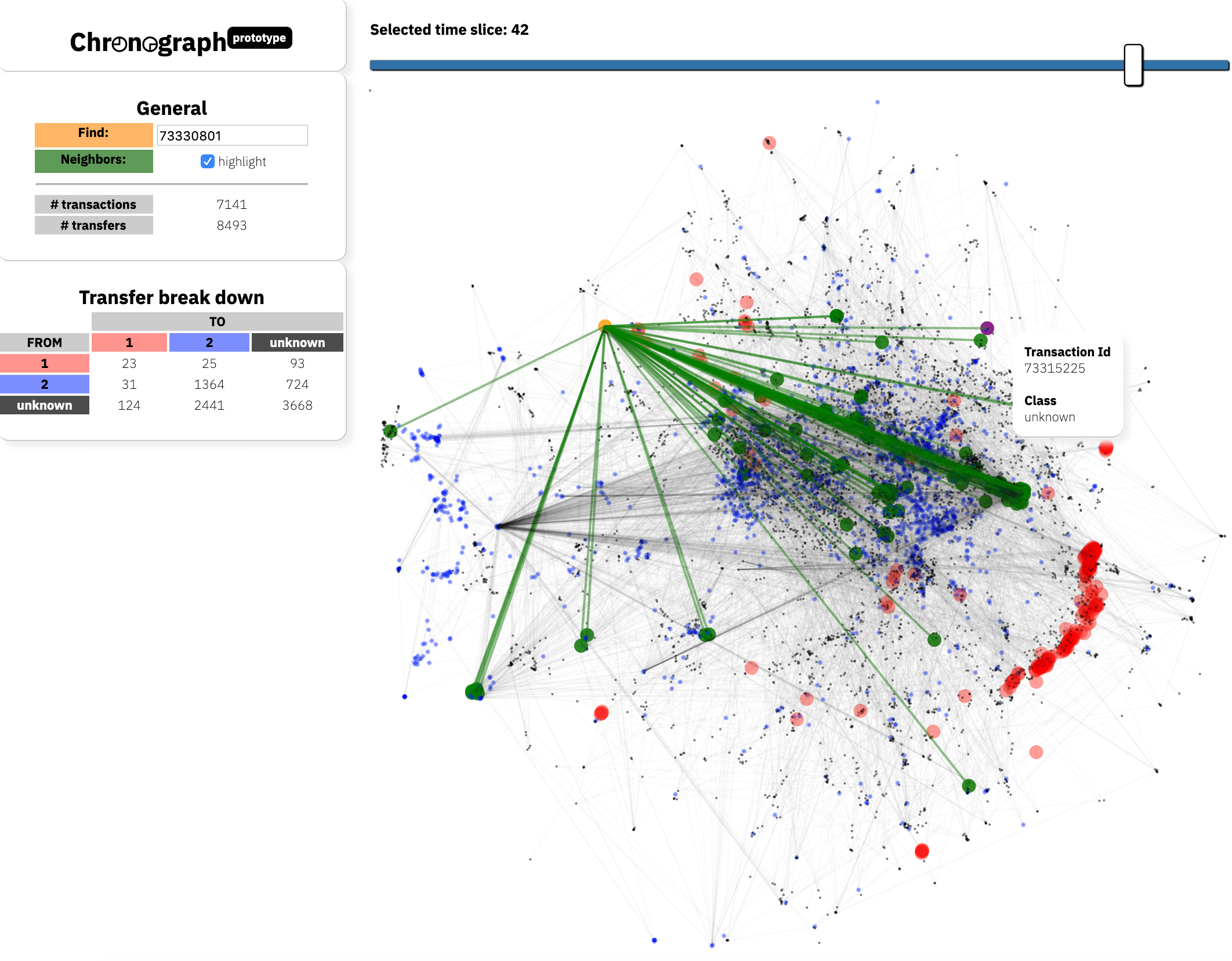}
  \caption{Chronograph User Interface: User can navigate through time-sliced transaction data and observe transaction patterns and patterns of change. Illicit transactions are dyed red. Further statistics are displayed on the left.}
  \Description{}
\end{figure}


\section{Summary}

In summary, we have set forth cryptocurrency forensics, and specifically Bitcoin, as a unique ecosystem for crowdsourcing the development of new methods to fight criminal activity. We have contributed a large, labelled transaction data set to the AML community, the likes of which has never before been publicly available. We have shared early experimental results using a variety of methods including Graph Convolutional Networks, and discussed possible next steps for algorithmic advances. We have provided a prototype for visualization of such data and models for augmenting human analysis and explainability. Most important, we hope to have inspired others to work on this societally important challenge of making our financial systems safer and more inclusive.

\begin{acks}
This work was funded by the MIT-IBM Watson AI Lab (mitibm.mit.edu), a joint research initiative between the Massachusetts Institute of Technology and IBM Research. Data and domain expertise were provided by Elliptic (www.elliptic.co).
\end{acks}

%
\bibliographystyle{ACM-Reference-Format}
\bibliography{bibliography}
%


\end{document}